\documentclass[]{article}
\usepackage{emulateapj,onecolfloat,psfig}

\newcommand{\eg}{{\it e.g.}}
\newcommand{\etal}{{\it et al.}}
\newcommand{\etc}{{\it etc.}}
\newcommand{\ie}{{\it i.e.}}
\newcommand\Lres{\hbox{$L_{\rm res}$}}
\newcommand{\HBWK}{Holley-Bockelmann, \etal}

\def\DF{{\sc DF}}
\def\pmb#1{\setbox0=\hbox{$#1$}%
  \kern-0.25em\copy0\kern-\wd0
  \kern.05em\copy0\kern-\wd0
  \kern-0.025em\raise.0433em\box0}
\def\spmb#1{\setbox1=\hbox{${\scriptstyle #1}$}%
  \kern-0.25em\copy1\kern-\wd1
  \kern.05em\copy1\kern-\wd1
  \kern-0.025em\raise.0433em\box1}
\def\bJ{\;\pmb{\mit J}}
\def\bm{\;\pmb{\mit m}}
\def\bw{\;\pmb{\mit w}}
\def\bOmega{\;\pmb{\Omega}}
\def\sbm{\;\spmb{\mit m}}

\slugcomment{\it To appear in ApJ}

\begin{document}

\twocolumn[
\title{Bar-halo Friction in Galaxies I: Scaling Laws}
\author{J. A. Sellwood}
\affil{Rutgers University, Department of Physics \& Astronomy, \\
       136 Frelinghuysen Road, Piscataway, NJ 08854-8019 \\ {\it
       sellwood@physics.rutgers.edu}}

\begin{abstract}
It has been known for some time that rotating bars in galaxies slow
due to dynamical friction against the halo.  However, recent attempts
to use this process to place constraints on the dark matter density in
galaxies and possibly also to drive dark matter out of the center have
been challenged.  This paper uses simplified numerical experiments to
clarify several aspects of the friction mechanism.  I explicitly
demonstrate the Chandrasekhar scaling of the friction force with bar
mass, halo density, and halo velocity dispersion.  I present direct
evidence that exchanges between the bar and halo orbits at major
resonances are responsible for friction and study both
individual orbits and the net changes at these resonances.  I also
show that friction alters the phase space density of particles in the
vicinity of a major resonance, which is the reason the magnitude of
the friction force depends on the prior evolution.  I demonstrate that
bar slow down can be captured correctly in simulations having modest
spatial resolution and practicable numbers of particles.  Subsequent
papers in this series delineate the dark matter density that can be
tolerated in halos of different density profiles.
\end{abstract}

]

\section{Introduction} 
Friction between a rotating bar and a massive halo was first reported
many years ago (Sellwood 1980) and has subsequently been worked on
sporadically (Tremaine \& Weinberg 1984; Weinberg 1985; Little \&
Carlberg 1991; Hernquist \& Weinberg 1992; Athanassoula 1996).  It
has, however, received a lot of attention in recent years as a
potential probe of, and structuring mechanism for, dark matter halos.
Debattista \& Sellwood (1998, 2000) argued that the fact that bars
appear not to have been slowed places an upper bound on the density of
the dark matter halo in barred disk galaxies.  Weinberg \& Katz (2002)
argue that the transfer of angular momentum from the bar to the halo
could reduce the central density of the dark matter halo by a
substantial factor.

Both of these claims have subsequently been disputed; Valenzuela \&
Klypin (2003) claim a counter-example of a bar that does not
experience much friction in a ``cosmo\-logically-motivated'' halo and
Holley-Bockelmann \& Weinberg (2005) find that cleverly-constructed
uniform-density halos can also avoid friction.  The argument by
Athanassoula (2003) that weak bars experience little friction is
clearly correct, but irrelevant for strong bars.  Only modest
reductions in halo density have been achieved so far
(Holley-Bockelmann, Weinberg \& Katz 2003) and Sellwood (2003) found
that contraction of the disk mass distribution as it lost angular
momentum to the halo dragged halo mass in with it, causing the density
to rise, rather than to decrease.

Valenzuela \& Klypin (2003) and \HBWK\ blame discrepant conclusions on
inadequacies respectively of the codes and of the number of particles
employed, but it is also likely that a good part of the differences
reported by these authors arise because the physical models differ.
There have been a few tests with different codes from the same initial
conditions and the results compare quite well (\eg\ O'Neill \&
Dubinski 2003; Sellwood 2003); other comparisons are reported here and
in Papers II \& III (Sellwood \& Debattista 2005a,b).

The present paper attempts to clarify the physics of dynamical
friction between a bar and a halo and to show that it can be correctly
reproduced in simulations of a size that is readily accessible with
current computational resources.  The counter-example reported by
Valenzuela \& Klypin (2003) is addressed in Sellwood \& Debattista
(2005c) and Paper II.  The constraint on halo density that can be
deduced from the existence of strong, fast bars and criticisms by
Athanassoula (2003) are addressed in Paper III.

\section{Theoretical background}
\label{theory}
Dynamical friction (Chandrasekhar 1943) is the retarding force
experienced by a massive perturber moving through a background of
low-mass particles.  It arises, even in a perfectly collisionless
system, from the vector sum of the impulses the perturber receives
from the particles as they are deflected by its gravitational field.
Equivalently, friction can be viewed as the gravitational attraction
on the perturber of the density excess, or wake, that develops behind
it as it moves, as was nicely illustrated by Mulder (1983).

\begin{figure}[t]
\plotone{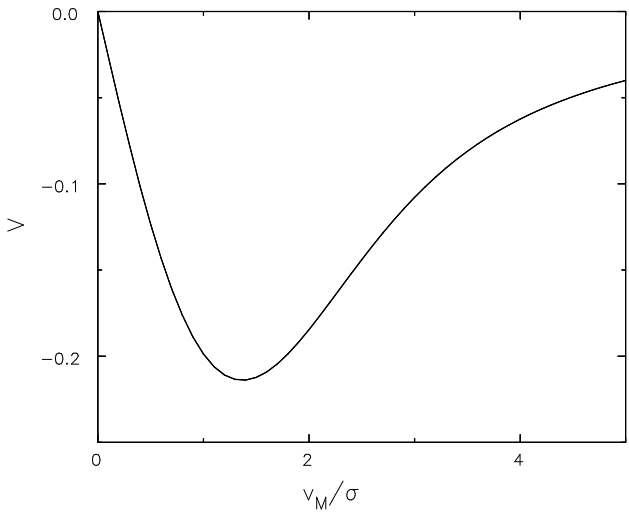}
\caption{\footnotesize The dimensionless acceleration function $V$
defined in equation (\ref{Chandra}) for the case of a Gaussian
distribution of velocities among the background particles.}
\label{dynv}
\end{figure}

Chandrasekhar's formula (eq.\ 7-17 of Binney \& Tremaine 1987;
hereafter BT) for the acceleration $a_M$ of a perturber of mass $M$
moving at speed $v_M$ through a uniform background, density $\rho$, of
non-interacting particles having an isotropic velocity distribution
with a 1-D rms velocity spread $\sigma$, may be written as
\begin{equation} 
a_M(v_M) = 4\pi \ln\Lambda G^2 {M \rho \over \sigma^2} V\left( {v_M
\over \sigma} \right).
\label{Chandra}
\end{equation}
Here, $\Lambda \;(= b_{\rm max}/b_{\rm min})$ is the argument of the
usual Coulomb logarithm, and the dimensionless function $V$ is drawn
in Figure \ref{dynv} for a Gaussian distribution of velocities; other
velocity distributions would yield a different functional form.

The simplifying assumptions in its derivation strictly invalidate
application of equation (\ref{Chandra}) to the physically more
interesting problem of friction in a non-uniform medium in which the
background particles are confined by a potential well and interact
with the perturber repeatedly.  Repeated encounters between the
perturber and the background particles require a more sophisticated
treatment.

\subsection{Orbits and resonances}
The motion of a particle pursuing a regular orbit in a smooth
ellipsoidal potential is triply periodic, and is most conveniently
described by action-angle variables (see BT).\footnote{Irregular
orbits that do not preserve three integrals of motion may be important
in real galaxies, but most discussion of dynamical friction has
focused on simple (usually spherical) potentials in which all
unperturbed orbits are regular.}  The three angular momentum-like
actions, $\bJ$, describe the amount of ``round and round'', ``in and
out'' and ``up and down'' motion associated with the orbit, while the
angles, $\bw$, specify the phases of these three separate oscillations
that underlie the motion.  One of the many advantages of these
variables is that the angles are defined such that they increase at a
uniform rate $\dot{\bw} \equiv \bOmega(\bJ)$ for each orbit.

In this paper, as in most other discussions, we will be concerned with
mild perturbations to a spherical potential, where the unperturbed
motion of any particle is confined to a plane.  In this case, there
are just two non-zero actions: the azimuthal action $J_\phi \equiv L$,
the total specific angular momentum, and the radial action, $J_r$.
The motion is then simply doubly periodic at angular rates
$\Omega_\phi$ and $\Omega_r$.

The angular frequency, $\Omega_\phi$, of a particle's mean motion
about the center is, in general, incommensurable with its angular
frequency of radial motion, $\Omega_r$.  However, all orbits can
appear to close when viewed from many different rotating frames.  An
observer rotating at the rate
\begin{equation}
\Omega^\prime = \Omega_\phi + k\Omega_r/m,
\end{equation}
would see the orbit close after $m$ radial oscillations and $k$ turns
about the center; Kalnajs (1977) draws an orbit in several of the most
important frames.

The angular momentum vectors of orbits in a spherical system are
distributed over all angles.  We define some direction as the $z$-axis
and use $\theta$ for the angle between that axis and the plane of each
unperturbed orbit, with $\theta$ having the same sign as the
$z$-component of the angular momentum $L_z = L\sin\theta$.  The
frequencies of the motion projected into the equatorial plane, $\theta
= \pi/2$, are independent of $\theta$, save only that the sign of
$\Omega_\phi$ is that of $L_z$.

We are interested in an ellipsoidal bar-like perturbation that rotates
about its minor axis, which we take to be the $z$-axis.  If the bar
rotates at an angular speed $\Omega_b$, there are many possible
resonances between the orbits of the particles and the perturbing
potential.  Resonances occur where $m\Omega_b = n\Omega_\phi +
k\Omega_r$, where $(m, k, n) \equiv \bm$ are three integers; $m$ is
even for a bar, $n=0,\pm m$, and $-1 \leq k \leq \infty$ (Tremaine \&
Weinberg 1984).  Three of the most important resonances are familiar
from disk dynamics: the corotation resonance (CR) where $n=m$ and
$k=0$, the inner Lindblad resonance (ILR) where $n=m=2$ and $k=-1$,
and the outer Lindblad resonance (OLR) where $n=m=2$ and $k=1$.
Negative values of $n$ lead to resonances with retrograde particles,
which seem to be of little dynamical importance.  Weinberg \& Katz
(2005; hereafter WK05) stress the importance a fourth: the direct
radial resonance (DRR) for which $n=0$ and $k=1$.  This last resonance
is absent for orbits in the bar plane and is most important for near
polar orbits.

\subsection{Friction in spheroidal systems}
Tremaine \& Weinberg (1984, hereafter TW84) developed the basic theory
of dynamical friction by a generic perturber in spherical systems.
Following the precepts of Lynden-Bell \& Kalnajs (1972), they derived
an expression for the ``LBK torque'' in a spherical system, which they
showed arises purely at resonances.  Weinberg (1985) gave a specific
evaluation of the LBK torque for the case of a bar.  The role of
resonances has been re-emphasized in recent work (Weinberg \& Katz
2002; Athanassoula 2003; \HBWK\ 2003; WK05).

As for the classical Chandrasekhar problem, friction can be viewed in
two equivalent ways: it can be seen either as arising from the
gravitational coupling between the bar and a misaligned density
response in the halo, or it can be interpreted as the net effect of
exchanges at resonances between the particles and the perturber.  The
misalignment can be understood in terms of interactions at resonances,
where halo particles gain or lose angular momentum.

The daunting expression for the LBK torque derived by TW84 (their
eq.~65) does not appear to resemble eq.~(\ref{Chandra}) above.
However, TW84 also reformulate the original Chandrasekhar problem in a
manner that highlights the similarities with the LBK torque in a
spherical system, but which contains no resonant terms because the
system is infinite.  A na\"\i ve guess, therefore, is that the LBK
torque causes the angular acceleration of the perturber to scale as
\begin{equation}
\dot\Omega_p \propto {M_p \rho_s \over \sigma_s^2} \Theta\left(
{\Omega_p a \over \sigma_s} \right),
\label{LBK}
\end{equation}
where $M_p$ is the mass of the perturber, $\rho_s$ and $\sigma_s$ are
the characteristic density and velocity dispersion of the spherical
system, and the dimensionless function $\Theta$ contains all the
complicated dependence on the details of the distribution function,
potential well, resonances, bar shape, \etc\ ~A characteristic length
scale $a$ is included in order to make the argument of $\Theta$
dimensionless.  The functions $\Theta(x)$ and $V(x)$ share the general
properties that they are negative, at least for isotropic distribution
functions (TW84), they must $\rightarrow 0$ as $x \rightarrow \infty$,
and that they can reasonably be expected to be $\propto x$ as $x
\rightarrow 0$.

Weinberg (2004) points out that the LBK torque formula derived by TW84
is not the full story, because their expression is evaluated for a
perturbation of fixed amplitude and pattern speed.  Weinberg finds
that the torque depends strongly on the previous time-dependence of
the perturbation, which he needed to take into account in order to
reconcile his perturbation theory with results from his simulations.
This improvement to the theory is a major step forward, as we show
here.

\subsection{Motivation}
The numerical experiments of Lin \& Tremaine (1983) showed that a
satellite orbiting a spherical system composed of much lighter
particles confined in a potential well experiences a frictional force
that scales very much as predicted by eq.~(\ref{Chandra}), or
eq.~(\ref{LBK}), despite the complications caused by resonances.
This, and other evidence, led BT to conclude that ``Chandrasekhar's
formula often provides a remarkably accurate description of the drag
experienced by a body orbiting in a stellar system''.

We might hope that the Chandrasekhar scaling would also hold for bars,
but there is no detailed check in the literature.  Weinberg (1985)
reports a few crude simulations of the Lin \& Tremaine type in support
of his theoretical calculations and he remarks (Weinberg 2004) that he
has recently made more such tests.

In \S\S\ref{restrict}\&\ref{scalings} I present a much more extensive
study using the Lin \& Tremaine technique for this slightly different
physical problem, and show that the expected scaling of
eq.~(\ref{LBK}) holds quite well.  \S\ref{orbits} reports a more
detailed study of resonant exchanges both of indiviual orbits and of
the ensemble of particles.  \S\ref{tdep} confirms Weinberg's (2004)
finding that the time dependence of the perturbation is of major
importance, and offers an explanation.  I present convergence tests
without self-gravity in \S\ref{converg} and I include the effects of
self-gravity between the halo particles in \S\ref{selfg} in order to
determine empirically the numbers of particles needed to reproduce the
correct frictional force in fully self-consistent simulations.

\section{Models and method}
\label{models}
\subsection{Halo models}
I create an $N$-body realization of a spherical mass distribution,
which for brevity I describe as a halo, although it could be any
spherical system of collisionless particles.  In \S\ref{restrict}, the
particles move in the smooth analytic gravitational potential of the
adopted halo and I neglect any interaction forces between the
particles.  Since I draw the particles from a distribution function
(\DF) that generates the adopted halo density, the particle
distribution is in equilibrium and does not evolve in the absence of
an external perturbation.  The isotropic halos start with no net
angular momentum.

I use three quite different halo models.  The first is a Hernquist
(1990) model, which has the density profile
\begin{equation}
\rho_{\rm H}(r) = {M_h r_{\rm H} \over 2\pi r(r_{\rm H} + r)^3}, 
\label{Hernquist}
\end{equation}
with total mass $M_h$.  The density profile declines as $r^{-1}$ for
$r \ll r_{\rm H}$ and as $r^{-4}$ for $r \gg r_{\rm H}$.  It should be
noted that this model differs only slightly from the NFW profile (see
Appendix), which appeared to be a reasonable fit to early simulations
(Navarro, Frenk \& White 1997) of the collapse of dark matter halos.
Hernquist gives the expression for the isotropic \DF\ that generates
the halo of eq.~(\ref{Hernquist}).  I do not employ the infinite
model, but remove from the \DF\ any particle with sufficient energy to
reach $r > 20r_{\rm H}$, so that the active mass in particles is $\sim
0.86M_h$ while they move in the analytic potential of the untruncated
halo.  The active density profile is very little affected for $r <
15r_{\rm H}$, while the bars I employ are typically much smaller, with
semi-major axes $\sim r_{\rm H}$.

I have also employed the well-known Plummer sphere
\begin{equation}
\rho_{\rm P}(r) = {3M_h \over 4\pi r_{\rm P}^3} \left( 1 + {r^2 \over
r_{\rm P}^2}\right)^{-5/2},
\label{Plummer}
\end{equation}
which has a uniform density core.  The isotropic \DF\ that generates
this density profile is a polytrope of index $n=5$ (BT, equation
4-104).  Again I eliminate any particle with sufficient energy to
reach $r > 20r_{\rm P}$, which is less than 1\% of the mass.

As a third halo model, I have adopted the singular isothermal sphere
(SIS), which formally has the scale-free density profile
\begin{equation}
\rho_{\rm I}(r) = {V_0^2 \over 4\pi Gr^2}.
\label{SIS}
\end{equation}
An isotropic \DF\ that generates this density has a 1-D velocity
dispersion $\sigma = V_0/ \sqrt{2}$.  The particles move in the exact
logarithmic potential of the untruncated sphere, but I generate active
particles from this \DF\ with a limited range of energies.  The upper
bound is set so that no particle has enough energy to pass an outer
radius $r_{\rm max}$, while the lower bound eliminates any particle
that would be bound inside some small radius $r_{\rm min}$.  I choose
$r_{\rm max} = 20a$ and $r_{\rm min} = 0.01 a$, where $a$ is my
adopted bar semi-major axis.  For this model only, I set the mass of
each halo particle proportional to $L^{1/2}$, where $L$ is its total
angular momentum, in order to obtain a disproportionately higher
density of particles in the inner parts of the halo.

\subsection{Bar}
The bar model is a homogeneous ellipsoid, which has the density
distribution
\begin{equation}
\rho_b = \cases{\displaystyle {3M_b \over 4\pi abc} & $\mu^2 \leq 1 $
\cr 0 & $\mu^2 > 1$ \cr}
\end{equation}
where $M_b$ is the mass of the ellipsoid, $a$, $b$ \& $c$ are its
semi-axes, with $a \ge b \ge c$, and
\begin{equation}
\mu^2 = {x^2 \over a^2} + {y^2 \over b^2} + {x^2 \over c^2}.
\end{equation}
I do not use the full field of this bar but employ only
non-axisymmetric parts of the field, excluding the monopole term in
order not to disturb the radial profile of the halo as I introduce the
bar.  This approach is similar to that first adopted by Hernquist \&
Weinberg (1992), who employed only an approximation to the quadrupole
term of the bar field.  Here, I determine the precise field of the bar
using a multipole expansion (\eg\ BT, \S2.4), but use only the
non-axisymmetric quadrupole ($l=2,\;m=2$) term, and in some cases
higher terms, to accelerate the halo particles.  (The odd-$l$ and
odd-$m$ terms all vanish because the bar is respectively symmetric
about the mid-plane and has 2-fold rotational symmetry, while the
$(l,0)$ terms do not rotate.  Because the bar lies in the equatorial
plane, terms with $l=m$ will be much larger than those with $l>m>0$.)
The internal and external contributions to the bar field over the
radial range of the bar are tabulated only once and stored; it is
straightforward to rotate the tabulated values through any desired
angle at each step.

The bar rotates at angular frequency $\Omega_b$ about its shortest
axis, and I introduce the bar smoothly by increasing $M_b$ as a cubic
function of time from zero at $t=0$ to its final value at $t=t_g$.  I
consider the model bar to be rigid, and to spin down due to loss of
angular momentum according to its moment of inertia about the shortest
axis, which is
\begin{equation}
I = {M_b \over 5} (a^2 + b^2).
\end{equation}

A rigid bar is essential for this study, even though it is unrealistic
in many ways.  I must make arbitrary choices for the bar mass, length,
axis ratio, density profile, and initial pattern speed, rather than
have all these parameters set by the dynamics of a disk.  In addition,
a rigid bar will not have the same moment of inertia as a figure of
the same density profile composed of active particles, which has a
pattern speed set by the mean precession rate of the bar particle
orbits.  Furthermore, the density profile of the bar should change as
the bar loses angular momentum and the mean radius of the bar
particles decreases, and a bar in an active disk may also grow by
trapping extra orbits.  The present study, however, requires rigid
bars with controlled parameters in order to determine how friction
depends on the bar properties.

\subsection{Numerical procedure}
The halo particles move in the combined fields of the halo and of some
non-axisymmetric terms of the rotating bar field.  The halo field is
rigid in the experiments described in \S\ref{restrict}, but some
self-gravity terms are included in \S\ref{selfg}.  I sum the
$z$-component of the net torque on the particles and use the negative
of this sum as the torque acting to accelerate the bar at every step,
so that the combined angular momentum of the halo and bar is
conserved.

My system of units is such that $G=M_h=r_x=1$, where $r_x = r_{\rm H}$
for the Hernquist halo and $r_x = r_{\rm P}$ for the Plummer halo.
For the SIS models, on the other hand, I choose $G=V_0=a=1$, for which
my unit of mass is $aV_0^2/G$.  Unless otherwise stated, the bar axes
are $a:b:c = 1:0.5:0.05$ and $t_g = 10$ in these units.

As particles in these models have a wide range of frequencies, I
divide the computation volume into a number of spherical zones
(typically 5) in which particles move with different time steps that
differ by factors of 2.  The integration step changes when particles
cross zone boundaries.  I have verified that the results presented
here do not depend on this scheme, or on the radii of the zone
boundaries, or on the adopted fundamental time step, neither are they
affected when the bar quadrupole field is replaced by the smooth
function adopted by Hernquist \& Weinberg (1992).
 
\section{Exploration of parameter space}
\label{restrict}
\subsection{Scaling with angular speed}
\begin{figure}[t]
\epsscale{1.0}
\plotone{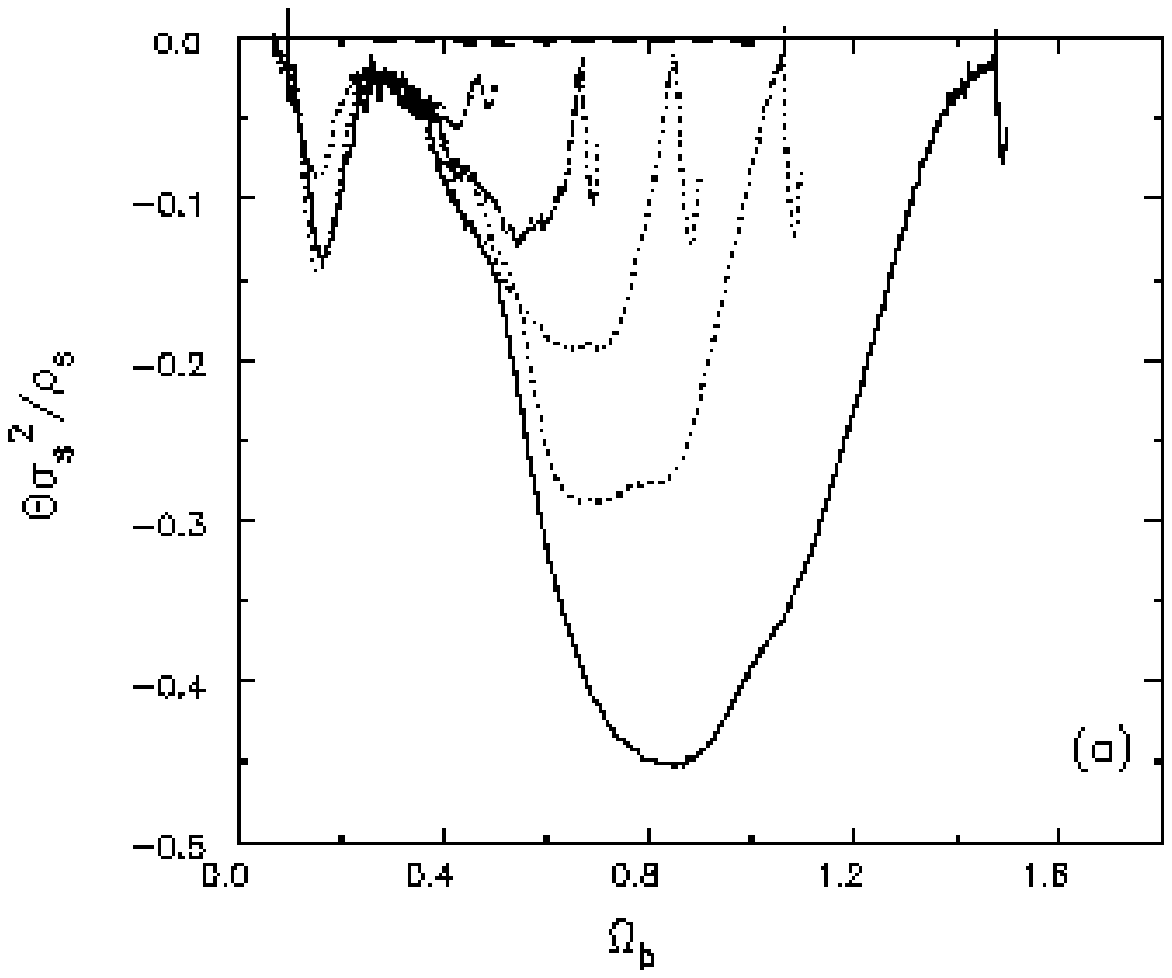}
\plotone{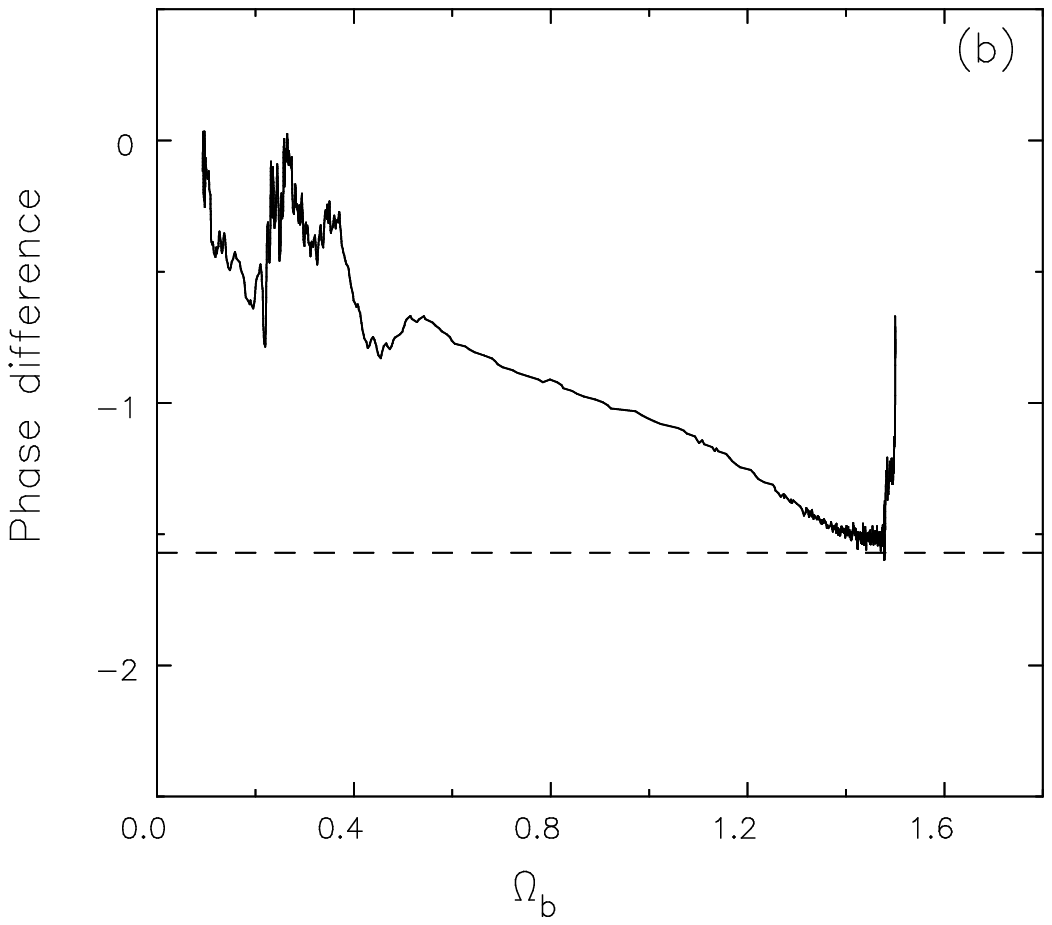}
\caption{\footnotesize (a) The variation of $\Theta$ with angular
speed of the bar.  The solid and dotted curves show the acceleration
in response to the quadrupole field of a fat (1:0.5:0.05) bar for
different initial $\Omega_b$ values.  The dashed curve shows the same
quantity when the perturbing force is the (4,4) component of the bar
field only.  (b) The phase lag in radians between the density response
in the inner halo and the bar from the run shown by the solid curve in
(a).  The response is approximately orthogonal to the bar direction
when $\Omega_b$ is large, and gradually shifts into alignment as the
bar slows.}
\label{vscale}
\end{figure}

The solid curve in Figure~\ref{vscale}(a) shows the function $\Theta$
for a Hernquist halo and a bar with axis ratios $a:b:c = 1:0.5:0.1$
and $a=r_{\rm H}$, when $\Omega_b = 1.5$ initially.  (The bar mass
$M_b = 0.01M_h$ and the halo is represented by 10M particles.)  Over
most of the range, the drag force on the bar varies quite smoothly,
peaking when $\Omega_b \simeq 0.8$.  An initial transient, associated
with the turn-on of the bar, is evident, and the curve also has a
feature near $\Omega_b \sim 0.2$.

In this case, the initial pattern speed of $\Omega_b=1.5$ is
unrealistically high, and corotation lies well inside the bar, at $r
\simeq 0.274a$.  The peak deceleration of the bar occurs when
corotation lies at the still unrealistically small radius $r \simeq
0.58a$.  The pattern speed must drop to $\Omega_b = 0.5$ to place
corotation at the end of the bar, by which time the drag force is
roughly one third of its peak value.

The dotted curves in Fig.~\ref{vscale}(a) show, for identically the
same bar and halo, the acceleration in a number of other experiments
with lower initial starting speeds.  The surprisingly large
differences confirm Weinberg's (2004) finding that the friction force
depends on the past evolution.  The largest differences arise at
angular speeds high enough for corotation to lie inside the bar and
differences are smaller when $\Omega_b \la 0.5$.  I discuss this
behavior more fully in \S\ref{tdep}.

Fig.~\ref{vscale}(b) shows the lag angle between the principal axes of
the bar and of the quadrupole response in the halo as a function of
$\Omega_b$ for the case when $\Omega_b=1.5$ initially.  I estimate the
position angle of the halo density response from the phase of the
(2,2) component of a high radial order spherical Bessel function (\eg\
Arfken 1985) transform of the particle distribution.  It is clear that
the response lags the bar by almost a right-angle at high $\Omega_b$,
and the lag angle generally decreases as the bar slows, becoming
nearly aligned with the bar when its rotation speed is very low.
Comparison with Fig.~\ref{vscale}(a) confirms, as it must, that the
torque is weak when the response is almost orthogonal to, or aligned
with, the bar and is greatest as the phase lag passes through
intermediate angles.  (Since the magnitude of the torque also depends
on the amplitude of the response, the maximum need not be when the
response is precisely $45^\circ$ out of phase, although it clearly
occurs close to this angle.)

The phase angle of the halo response shown in Fig.~\ref{vscale}(b) is
not hard to understand.  The following argument is specific to
spherical potentials, but it generalizes to aspherical cases.

As already noted, any unperturbed orbit will close in a frame that
rotates at the rate $\Omega_{\sbm} = (n\Omega_\phi + k\Omega_r)/m$
($n\neq0$), but only those orbits for which $\Omega_{\sbm} = \Omega_b$
also close in the frame of the rotating perturbation, and are exactly
in resonance.

\begin{figure}[t]
\epsscale{1.0}
\plotone{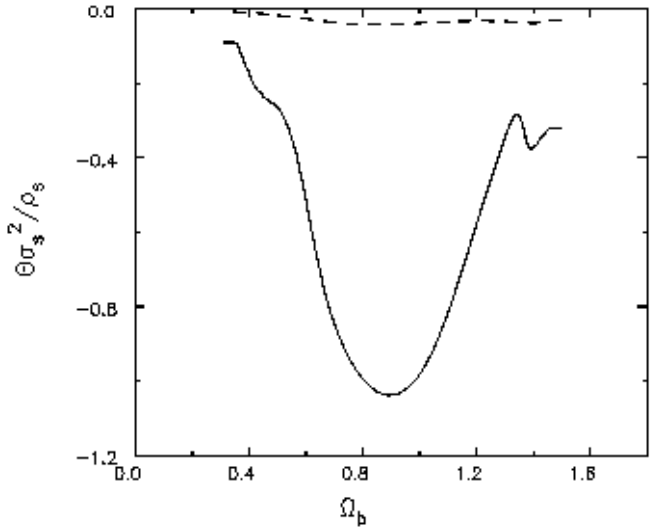}
\caption{\footnotesize As for Fig.~\ref{vscale}(a) but for a thinner
bar with axis ratios 1:0.2:0.05.  Note the different scales between
the two figures.}
\label{skinny}
\end{figure}

Most orbits are not resonant, however, and $\Omega_{\sbm} \neq
\Omega_b$ for any $\bm$.  For these general orbits, the perturbation
adds to its otherwise axisymmetric time-averaged density, a forced
non-axi\-symmetric distortion that corotates with the bar.  As happens
for simple harmonic oscillators, the driven response is in phase, or
aligned, with the bar when $\Omega_b < \Omega_{\sbm}$, and is
perpendicular to the bar for higher bar pattern speeds.  Thus the
forced response of an orbit switches from perpendicular to alignment
as the bar slows across its resonant frequency; the change of phase is
gradual because the resonance is broadened by the changing pattern
speed.  Since many orbits are present with a wide range of precession
rates for each resonance, the net response is the aggregate of many
orbits.

\subsection{Different strength bar}
Fig.~\ref{skinny} shows the much stronger frictional deceleration for
a thinner bar with axis ratios 1:0.2:0.05.  The quadrupole potential
of this bar peaks at about twice the value of the fatter bar used in
Fig.~\ref{vscale}(a), but at a smaller radius; the two fields could
not be matched by scaling, therefore.  Nevertheless, the curve has a
similar shape.

\begin{figure}[t]
\plotone{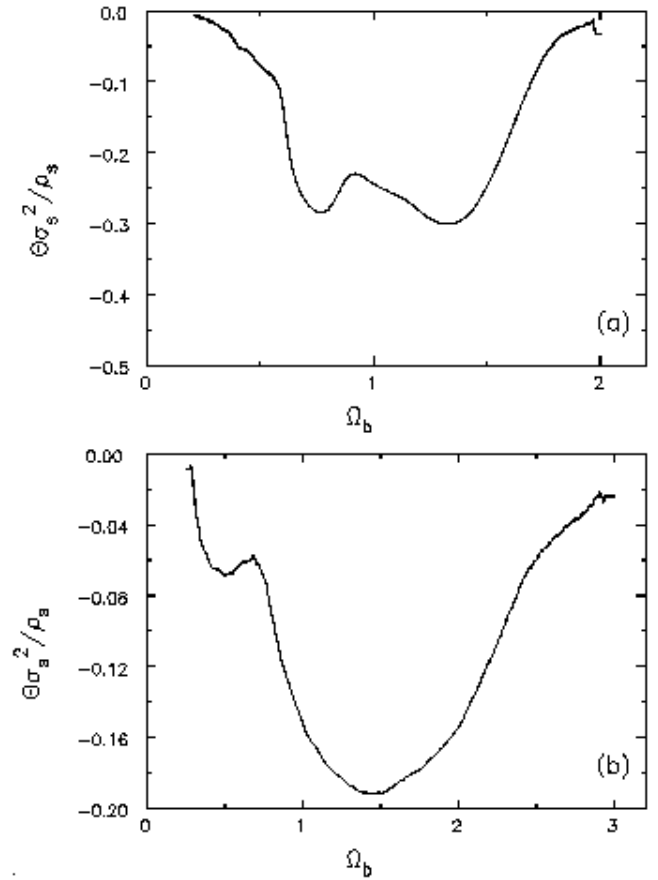}
\caption{\footnotesize As for Fig.~\ref{vscale}(a), but (a) for the
Plummer halo and (b) for the singular isothermal sphere.}
\label{msplum}
\end{figure}

\subsection{Four-fold and higher potential terms}
The dashed curves in Figs.~\ref{vscale}(a) \& \ref{skinny} show the
frictional deceleration for a 20 times more massive bar when only the
$l=4, \; m=4$ term of the bar field is used to force the halo.  Since
evolution with a 1\% mass bar would be impracticably slow, I used a
20\% mass bar in both cases, and scaled the curves appropriately.  The
relative strengths of the quadrupole and higher-order components of
the perturbing field depend on the bar shape; the (4,4) component of
the potential has a peak amplitude of $\sim 1/6$ that of the
quadrupole for the fatter bar, whereas it is a larger fraction ($\sim
1/3$) for the thinner bar.  Yet the contribution to friction from the
(4,4) term is small even for the sharper bar (Fig.~\ref{skinny});
higher order terms are still less important.

Since friction is dominated by only the lowest-order, non-axisymmetric
term of the potential, it is possible for even low-spatial resolution
codes, such as that used by Debattista \& Sellwood (1998, 2000), to
reproduce the frictional drag quite accurately.  Thus the suggestion
by Valenzuela \& Klypin (2003) that their different result was due to
inadequate spatial resolution in the earlier work is unlikely to be
correct.  The different results obtained in these two studies will be
discussed further in Paper II.

\subsection{Other halo density profiles}
Figure~\ref{msplum}(a) shows the variation of $\Theta$ with angular
speed for the Plummer halo, while Fig.~\ref{msplum}(b) is for the
singular isothermal sphere (SIS).  These are directly comparable with
Fig.~\ref{vscale}(a), which is for the Hernquist halo.  The angular
velocity dependence of the frictional acceleration is noticeably
different in both cases.

The bar in the Plummer halo has a semi-major axis $a=R_{\rm P}$ and a
mass $M_b = 0.02M_h$.  Friction has two approximately equal peaks when
$\Omega_b \simeq 1.4$ and $\Omega_b \simeq 0.7$, but the angular speed
of this bar must drop to $2^{-3/4}\simeq 0.6$ before corotation moves
outside the bar.

The behavior for the SIS is different again.  Corotation is at the end
of the bar when $\Omega_b = 1$, where friction is again past its peak.
Note that features can arise in this curve because the scale-free
nature of this model is broken in two ways: the perturbation has a
definite linear size and the energy range of the active particles is
restricted.  The mass of the bar in this case, $M_b = 0.2aV_0^2/G$.

These curves, and that in Fig.~\ref{vscale}(a), indicate the
functional form of $\Theta$ in eq.~(\ref{LBK}).  Since the bar is the
same in all three cases, the differences stem directly from
differences in the halo mass profiles, which affect the frequencies
and particle density at the different resonances.  Such differences
are the analog of having a different velocity distribution for the
background particles in the Chandrasekhar problem.
 
\begin{figure}[t]
\epsscale{1.0}
\plotone{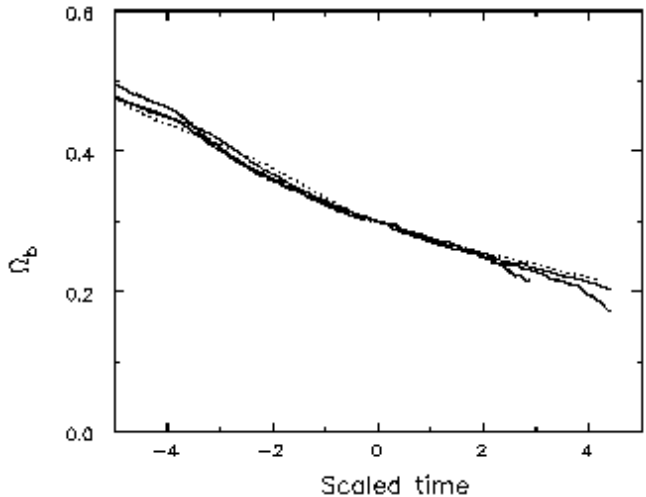}
\caption{\footnotesize The bar pattern speed as a function of
normalized time for a series of experiments with differing mass bars,
but all other properties held fixed.  Bar masses are 0.5\%, 1\%, \&
2\% (all solid lines) and 5\% (dotted line) of the halo mass.  The
times are scaled by the bar masses and shifted horizontally to
coincide when $\Omega_b = 0.3$.}
\label{mscale}
\end{figure}

\section{Scaling with other parameters}
\label{scalings}
In this section, I check directly whether the acceleration scales with
bar mass, halo density and halo velocity dispersion, as suggested by
eq.~(\ref{LBK}).  All experiments reported in this section employ 10M
particles, as were also used for those shown in Figs.~\ref{vscale},
\ref{skinny}, \& \ref{msplum}, but the initial bar pattern speed is
set at the more conventional value such that corotation is at the end
of the bar.  I illustrate these tests for the Hernquist halo; the
other halo models scale about as well, except for the test with
velocity dispersion, as discussed in \S\ref{dispscale}.

\subsection{Scaling with bar mass}
Figure \ref{mscale} shows the time evolution of the pattern speed for
bars of different masses, in the Hernquist halo; the bar pattern speed
is a smoother function than its acceleration.  Each line in this
Figure is from a run with a different bar mass in the range $0.005
\leq M_b/M_h \leq 0.05$ and $\Omega_b=0.5$ initially; the time axis is
scaled by $M_b$, and the curves have all been shifted horizontally to
coincide at the moment at which $\Omega_b=0.3$.  The curves overlay
almost perfectly, indicating that frictional acceleration scales
linearly with $M_b$ in this regime, as expected from eq.~(\ref{LBK}).

\begin{figure}[t]
\plotone{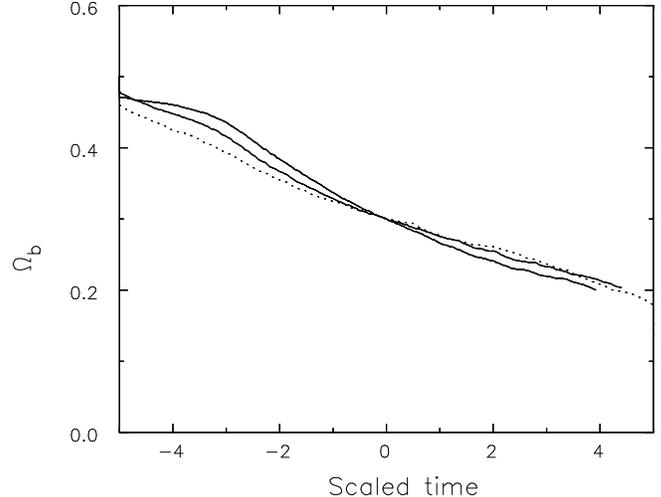}
\caption{\footnotesize As for Fig.~\ref{mscale}, but for the case when
the halo density is scaled.  The dotted curve is for the case when the
halo density is doubled.  See \S\ref{denscale} for details.}
\label{rhoscale}
\end{figure}

\subsection{Scaling with halo density}
\label{denscale}
Equation (\ref{Chandra}) predicts that the acceleration is
proportional to the density of the background.  In the present more
realistic case, the density of the background is a function of radius
and it might seem that I would have to evaluate the LBK torque for a
number of different halos to obtain a testable prediction.
Fortunately, there is a much simpler way to test for the density
dependence: I simply scale the density of the entire halo without
changing the potential well in which the particles move.  This ploy is
equivalent to treating some fraction of the halo as rigid if the
density is reduced but, since the density and potential of the halo
need not be self-consistent in these restricted experiments, it is
also possible to increase its density.  I adopt this admittedly
artificial strategy here simply to test for the expected linear
scaling.

Figure~\ref{rhoscale} shows the results with a 2\% mass bar, with
Hernquist halos that have 0.5, 1 and 2 times the density given in
eq.~(\ref{Hernquist}).  Again the time axis scale is proportional to
the density, and curves are shifted so that they coincide when
$\Omega_b=0.3$.  While not quite as convincing as Fig.~\ref{mscale},
the similarity of the curves shows that the expected linear scaling
holds approximately, in satisfactory agreement with the prediction of
eq.~(\ref{LBK}).

\begin{figure}[t]
\plotone{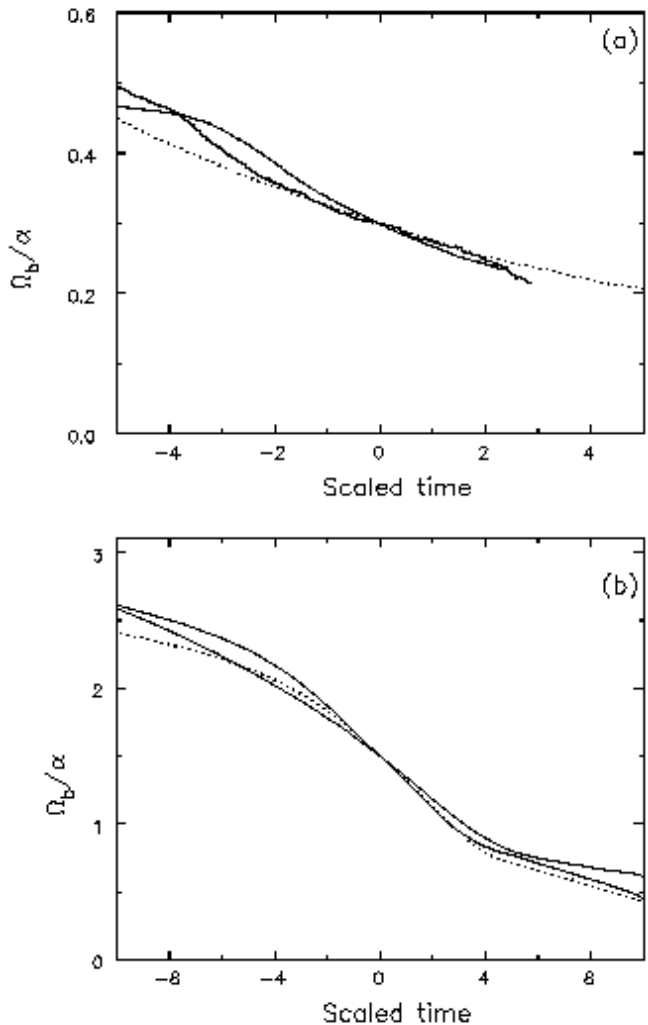}
\caption{\footnotesize (a) As for Fig.~\ref{mscale}, but for the case
when the halo velocity dispersion and potential are scaled.  The
dotted curve is for the largest velocity dispersion. (b) Same as (a),
but for the singular isothermal sphere.  See \S\ref{dispscale} for
details.}
\label{sigmascale}
\end{figure}

\subsection{Scaling with halo velocity dispersion}
\label{dispscale}
Finally, I test the scaling with velocity dispersion predicted by
eq.~(\ref{LBK}).  Again I employ a trick: by scaling the gravitational
potential by a factor, $\alpha^2$, and the velocity dispersion,
$\sigma$, by $\alpha$, we continue to have an equilibrium model with
an unchanged density profile.  (The halo density and potential are no
longer self-consistent, of course.)  For this series of tests,
however, I must also scale the bar pattern speed, since the argument
of $\Theta$ contains the dispersion, $\sigma$.

Figure~\ref{sigmascale}(a), which plots $\Omega_b/\alpha$ as a
function of $Mt\alpha^{-3}$, shows that when the velocity dispersion
is increased or decreased by a factor $\sqrt2$, the evolution of the
scaled bar angular speed does vary at approximately half or double the
rate, respectively, as predicted by eq.~(\ref{LBK}), but the curves do
not match up as well as in Figs.~\ref{mscale} \& \ref{rhoscale}.

Since scaling the potential well affects all the resonances, the
particle density at each resonance does not scale in a simple manner,
and formula (\ref{LBK}) is too na\"\i ve.  However, scaling should be
restored in a scale-free model.  The same scaling test with the SIS
supports this expectation; the curves in Fig.~\ref{sigmascale}(b)
follow each other more closely than in Fig.~\ref{sigmascale}(a), which
was for the Hernquist halo.

\subsection{Summary}
The present problem differs from that considered by Chandrasekhar in
many ways: the perturber is an extensive rigid body moving at changing
angular frequency through an inhomogeneous sea of particles, which
encounter the perturbation in a periodic fashion.  Yet, this series of
experiments has demonstrated that the deceleration of the bar caused
by dynamical friction from non-interacting halo particles scales with
bar mass and halo density as predicted by eq.~(\ref{LBK}), which so
closely resembles Chandrasekhar's formula (1).  The na\"\i ve scaling
with velocity dispersion holds approximately for a general halo, and
rather better for a self-similar halo, such as the SIS.  The scaling
with these parameters is an inevitable consequence of dynamical
friction being second order effect caused by the interaction between
the perturber and its own wake.

\section{Exchanges at Resonances}
\label{orbits}
In this section, I present a more detailed analysis of one experiment
in order to shed more light on resonant interactions, and the
time-dependence of the frictional acceleration.  The variation of
$\Theta(\Omega_b)$ for this ``fiducial'' experiment is shown by the
dotted line that starts with $\Omega_b=0.5$ in Fig.~\ref{vscale}(a) --
\ie\ the experiment with the lowest initial $\Omega_b$, and the only
one shown in this figure to start with corotation outside the bar.

\subsection{Adiabatic invariants}
The interaction between an orbit and a perturbing bar potential was
discussed by Lynden-Bell (1979) in the analogous case of disks.  In
that geometry, the unperturbed orbit of a particle in resonance closes
in the frame that rotates with the perturbation, and Lynden-Bell
pointed out that a star close to a resonance could be regarded as
pursuing a closed figure that precesses slowly relative to the major
axis of the perturbation.

When $m=n$, for the disk-like resonances, the unperturbed orbit
precesses at the rate $\Omega_{\sbm} \equiv (n\Omega_\phi +
k\Omega_r)/m$.  In general, $\Omega_{\sbm} \neq \Omega_b$, and the
difference, $\Omega_{\sbm} - \Omega_b = \Omega_s$ is a slow frequency
close to the resonance; \ie\ $|\Omega_s/\Omega_\phi| \ll 1$.  Since
the time taken for the star to complete an orbit round the closed
figure is short (one or two radial periods), the action $J_f \equiv
J_r - kJ_\phi/m$, associated with this fast motion will be
adiabatically invariant.  On the other hand, the slow rate of
precession of the orbit relative to the bar allows a large change to
the ``slow'' action.

The situation is more complicated for the DRR, since the orbit does
not close in the bar frame and the adiabatic invariant is the total
angular momentum, $L$ (WK05).  Conservation of $L$ might seemingly
preclude any angular momentum exchanges with the bar.  However, this
resonance is most important for highly inclined orbits, and while $L$
is conserved, the plane of the orbit, and therefore $L_z$, is changed
at the resonance; thus interactions at the DRR are able to slow the
bar.

Because even quite mild potential perturbations can produce non-linear
orbital responses at resonances in this manner, TW84 examined the
resonant terms more carefully.  They showed that second-order
perturbation theory remains valid provided that particles make a
``fast'' passage through the resonance.  By this, they mean that the
pattern speed of the bar is changing sufficiently rapidly that ``large
non-linear resonant perturbations do not have time to develop before
the star has crossed the resonance.''  In effect, the star crosses the
resonance in less than one libration period of the figure.  Since the
precession rate is slow, TW84 and Weinberg (1985) argued that fast
passage through the resonance is the correct assumption, and
second-order perturbation theory should be valid.

\begin{figure}[p]
\epsscale{.83}
\centerline{\psfig{figure=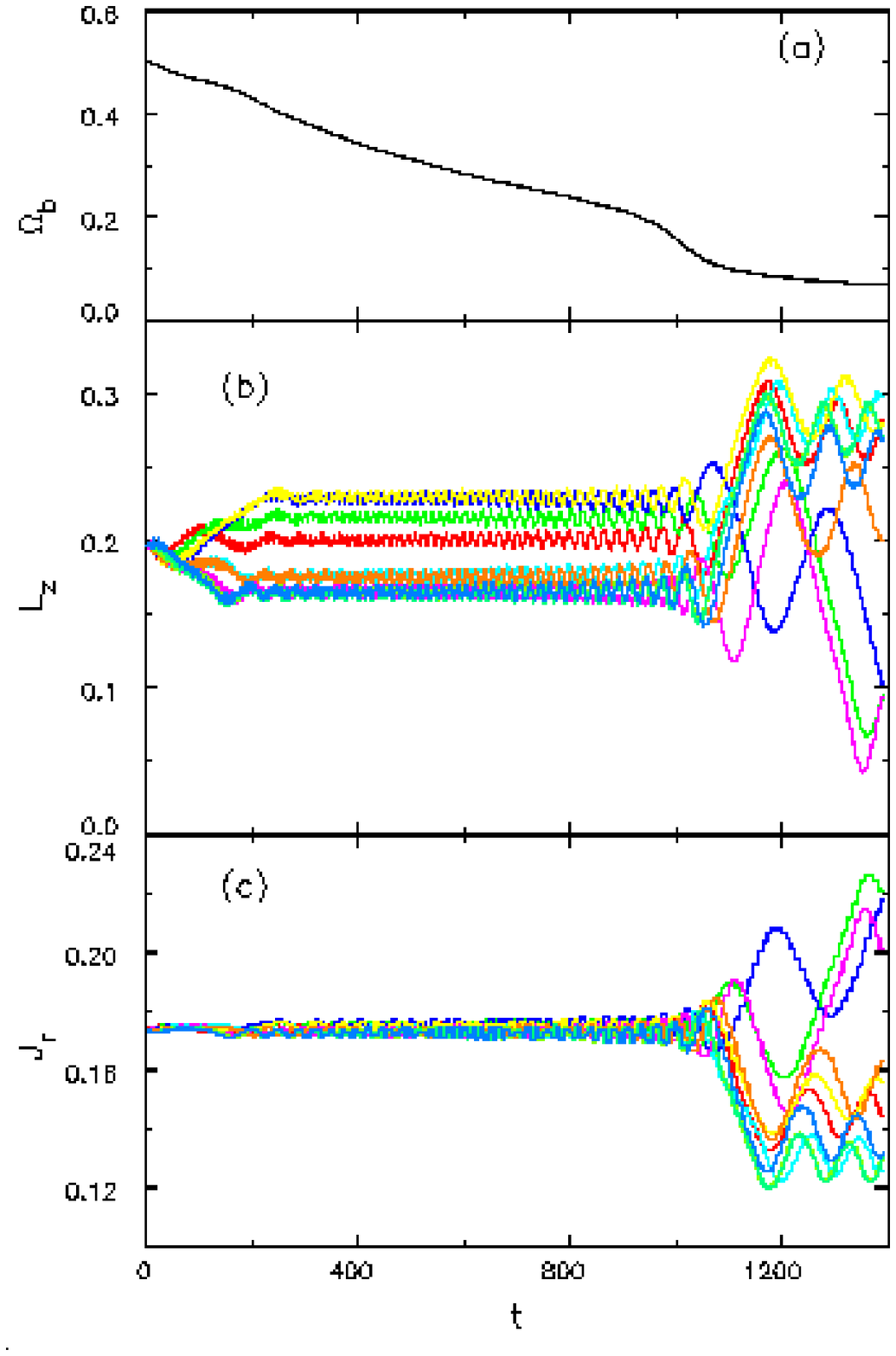,width=.83\hsize,angle=0,clip=}}
\centerline{\psfig{figure=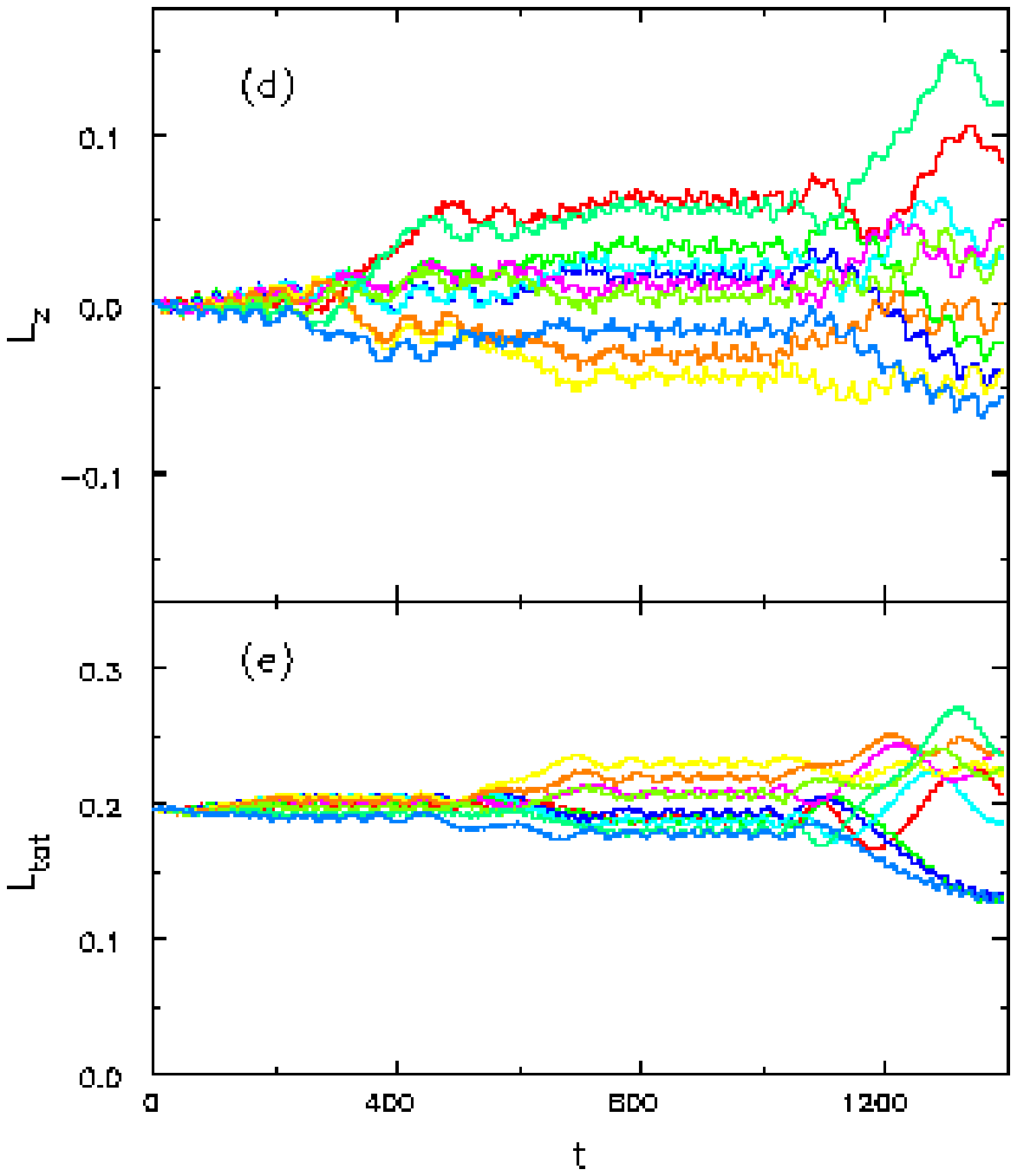,width=.83\hsize,angle=0,clip=}}
\caption{\footnotesize (a) The evolution of the pattern speed in the
fiducial run.  (b) The instantaneous angular momenta of ten particles
as a function of time.  The particles move in the mid-plane and all
start with the same energy and angular momentum but have equally
spaced phases.  The quasi-periodic oscillations on each line are the
forced response to the bar potential, the large changes, which are not
modulated for the most part, are caused by passages through
resonances. (c) An approximation to the radial action for the same
orbits shown in (b).  (d) The instantaneous $z$-component of angular
momentum of a different ten particles that start on exactly polar
orbits, but have the same energy and angular momentum as in (b) have
equally spaced phases.  (e) The total angular momentum of the
particles shown in (d).}
\label{fastf}
\end{figure}

\subsection{Individual orbits}
Figure \ref{fastf} presents a direct test of this assumption in the
fiducial run.  The time evolution of $\Omega_b$ is shown in panel (a),
while panel (b) shows the evolution of the instantaneous $L_z$ for ten
particles orbiting in the mid-plane.  The orbits shown are those of
test particles run in a reconstruction of the time evolution of the
total potential, which requires knowledge only of the bar phase at
every step in the original simulation.  All have the initial energy
$E=-0.38$ and angular momentum, $L_z = 0.4L_{\rm max}(E)$ but are
spaced equally in initial phases.  The short-period oscillations in
$L_z$ of each particle are forced as each particle moves in the bar
field.  Larger, and lasting, changes occur as particles pass through a
resonance.  Individual orbits can either gain or lose, depending on
their phase relative to the bar.

The group of particles selected is typical of the many hundreds I have
examined.  Large changes in $L_z$ occur over the time interval $t
\simeq 50$ to $t \simeq 200$ during which time $\Omega_b$ decreases
from $\sim 0.45$ to $\sim 0.4$.  About half the particles gain in
angular momentum by a up to $\sim 20\%$ as they are caught in a
corotation resonance, while the particles at other orbital phases lose
similar amounts.  These same particles also pass through an inner
Lindblad resonance around $t\sim 1100$ where a majority gain $L_z$.
Notice that friction is moderate when these particles pass through the
CR, but that the late second blip in friction (Fig.~\ref{vscale}a)
occurs as the particles pass through the ILR.  Many other in-plane
orbits that I have examined also show two separate periods of angular
momentum exchange with the bar.

It should be noted that the more tightly bound the orbit, the higher
its intrinsic frequencies.  Some of the most tightly bound orbits
indeed pass through the ILR at early times, as stressed by WK05.
However, these orbits are confined close to the center, where they
couple weakly to the inner part of the perturbation, whose amplitude
decays quadratically towards the center.  Their consequent modest
gains in angular momentum contribute little to the overall torque,
which is dominated by the CR and OLR exchanges by larger orbits, as I
show in \S\ref{gradients}.  The bar has already become uninterestingly
slow by the time the larger orbits cause strong friction at the ILR.

TW84 note that the frictional torque will be reduced, and the sign of
the exchange between the particle and the perturbation difficult to
predict, if the halo particles pass slowly through the resonance (as
defined above).  TW84 and Weinberg (1985) expected fast passage
through the resonance, but offered no proof.  Diagrams such as
Figure~\ref{fastf}(b) allow us to examine this key issue.  The changes
for each particle at CR are generally non-oscillatory, indicating that
the particle is not librating in the resonance, and therefore the
passage is fast.  The story at the ILR is less clear, however, as the
instantaneous $L_z$ of each particle continues to oscillate right to
the end.  It is likely that these oscillations are simply the forced
responses to the bar potential; their period seems to be about 100
time units, which is reasonable for precession of closed elliptical
figures with respect to the slowly-rotating bar.  These oscillations
aside, the larger changes in $L_z$ mostly, with perhaps a couple of
exceptions, seem to occur within a single oscillation period.

Thus Figure~\ref{fastf}(b), and hundreds of other orbits that I have
examined, show that most passages through resonance can indeed be
treated by the LBK approach.  Since I have not examined every corner
of phase space, this does not, of course, amount to a proof that the
fast passage assumption is adequate, but it does suggest it is good
for most orbits.

The adiabatic invariance of the fast action is also borne out, as
shown in Fig.~\ref{fastf}(c).  Determination of the exact radial and
azimuthal actions in a rotating and slowing, non-axisymmetric
potential would be very difficult, but approximate values are good
enough for our purpose here.  The instantaneous value of $L_z \simeq
J_\phi$, and we adopt $J_r \simeq 0.5a^2\pi/\tau$, where $a$ is the
semi-radial excursion of the particle and $\tau$, the radial
half-period, is determined by the time between the most recent
passages through peri- and apo-center for the particle.
(Properly-defined actions would not exhibit short-period oscillations,
and the non-smooth variations of $J_r$ at late times reflect a change
in the estimated value each time the radial direction of the particle
reverses.)

At the CR, $k=0$ and therefore $J_f=J_r$; it can be seen that the
approximate radial action is almost unchanged during the changes to
$L_z$ around $t\sim100$.  (Note the difference in scales between
panels b and c.)  The opposite is true for the changes around
$t=1100$, where it is clear that particles that gain $L_z$ also lose
$J_r$ and {\it vice-versa}, as predicted by conservation of $J_f$ when
$k=-1$ for the ILR.  In fact, $\Delta J_r \simeq -\Delta L_z/2$ in
these cases, confirming that resonance responsible for these large
changes is indeed the ILR.

Fig.~\ref{fastf}(d) and (e) present a similar study of orbits that
start with the same total angular momentum and energy as in (b) and
(c), but which are initially oriented perpendicular to the bar plane
so that $L_z=0$ for all initially.  These orbits pass through the DRR
between $300 \la t \la 400$, and it can be seen that some gain $L_z$
and others lose.  However, panel (e) shows that the total $L$ of these
orbits does not change during this period, confirming that this
resonance merely re-orients the plane of each orbit.  (These same
particles suffer changes to $L$ at later times when they pass
through other resonances.)

\begin{figure*}[t]
\epsscale{1.5}\plotone{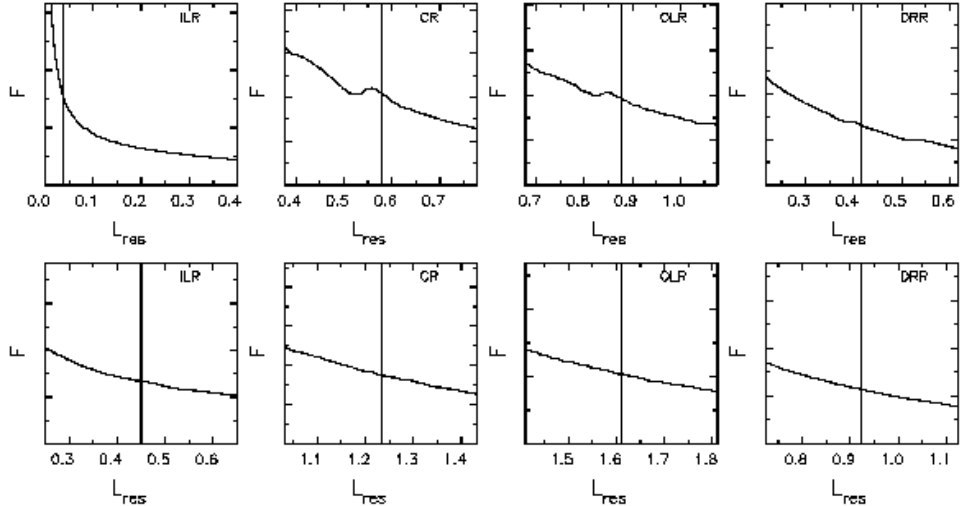}\epsscale{1.0}
\caption{\footnotesize The function $F(\Lres)$ as defined in the text
at the most important resonances at two different times in the
fiducial simulation; the top row shows curves at $t=200$, while the
bottom row is for $t=1000$.  The vertical line shows location of the
resonance at the moment illustrated, and the range of abscissae is set
to show a region around the resonance containing $\sim 10$\% of the
halo particles.  The vertical scale is linear from zero, but is
otherwise arbitrary.}
\label{Lsmooth}
\end{figure*}

\subsection{Integrated changes}
\label{gradients}
It is interesting also to examine the changes to the distribution of
particles about the main resonances as friction proceeds.  Extracting
this information from simulations is not completely straightforward
because the angular momentum of a resonant particle on a nearly
circular orbit differs from that of another particle at the same
resonance that has a highly eccentric orbit.  \HBWK\ (2003) and WK05
draw contours of the net changes of angular momentum over a short time
interval in $(E,L/L_{\rm max})$ space, where $L_{\rm max}(E)$ is the
specific angular momentum of a circular orbit of specific energy $E$.
In order to suppress shot noise, these figures require considerable
smoothing, even when very large numbers of particles are employed.

Perturbation theory (TW84, Weinberg 2004) indicates that friction
depends on gradients of the \DF\ w.r.t.\ both $J_r$ and $J_\phi$,
although the coefficient of the radial action derivative is $k$, which
is zero at CR.  Since $J_f \equiv J_r - kJ_\phi/m$, is adiabatically
invariant, changes in $J_r$ must be related to changes to $J_\phi$,
and we need examine the distribution with respect to a single action
only.  The natural choice is $L_z$, the quantity of greatest interest
for friction.

At any given resonance, the particles that undergo the largest changes
have $\Omega_{\sbm} \simeq \Omega_b$.  We define $f(J_r,J_\phi)$ to be
the density of particles in the space of these two
actions,\footnote{This is not the distribution function as usually
defined, but is the \DF\ integrated over all orbit phases and the
inclination angle of the orbit plane $\theta$.} and determine the mean
value of $f$ at fixed $\Omega_{\sbm}$.  The mean value is defined by
the path integral
\begin{equation}
F(\Omega_{\sbm}) = {\int_C f(J_r,J_\phi) \, dl \over \int_C  dl},
\label{FLres}
\end{equation}
with the path $C$ being the locus of constant $\Omega_{\sbm}$ through
$(J_r,J_\phi)$ space from $J_r=0$ to $J_\phi=0$.  In practice, the
numerator is a kernel estimate from the values of $\Omega_{\sbm}$
computed from all the particles at each separate resonance.  Since
$\Omega_{\sbm}$ is a single-valued function of the angular momentum,
$\Lres$, of a circular orbit, I use this as the abscissa
\begin{equation}
F(\Lres) = {\sum_i w(\Lres - L_{{\rm res},i}) \over \int_C  dl},
\end{equation}
where $L_{{\rm res},i}$ is the angular momentum of a circular orbit
that has the same $\Omega_{\sbm}$ as the $i$-th particle, $w$ is a
kernel function, and the denominator is unchanged.

The average defined in eq.~(\ref{FLres}) is not a canonical
transformation, and therefore does not preserve phase space volume.
Thus we cannot expect the gradient of $F(\Lres)$ to be a completely
reliable surrogate for the gradients in the \DF\ -- a Jacobian factor
may alter the slope.  But since $F$ is defined as the average of $f$
at fixed frequency difference from the resonance, we should expect its
slope to give a general indication.

Figure~\ref{Lsmooth} shows the function $F(\Lres)$ at the ILR, CR, OLR
and DRR at two separate times during the evolution of the fiducial
model.  A significant fraction of the particles in the simulation, of
order 10\%, contributes to the curve in each panel implying a rather
low-level of shot noise.

The strongest feature is for CR at $t=200$ (middle panel of top row),
where $F$ shows a clear peak just to the low-$L$ side of the
resonance, which is marked by the vertical line.  Near to the
resonance, particles cross corotation in both directions on horse-shoe
orbits (Binney \& Tremaine 1987, \S7.5) as they gain and lose $L_z$.
The negative gradient in $F$ implies there are more particles on the
low-$L$ side, and therefore an excess of gainers over losers.  This
imbalance causes a net gain of angular momentum by the particles, and
therefore a net loss by the bar -- producing the required friction.

If the pattern speed were to stay constant, the imbalance would tend
to flatten the slope of $F$, and the distribution of particles about
the resonance would approach kinetic equilibrium in which there would
be more nearly equal numbers of particles crossing in both directions.
But as $\Omega_p$ declines, the resonance keeps moving to larger
$\Lres$ (frequency is decreasing function of angular momentum), and
equilibrium is never established; instead, the density of particles
about the dominant resonance(s) responsible for friction maintains a
shoulder, or excess of particles, on the low-$L$ side of the
resonance, as shown in Fig.~\ref{Lsmooth}.

The pronounced feature at CR at $t=200$ is characteristic of the
distribution over most of the evolution, and changes at CR dominate
the torque.  A similar, but less pronounced, feature can be seen at
the OLR at $t=200$, but its importance fades as the bar slows.  The
DRR also shows a definite, though weaker, feature at $t=200$; note
that the range of $\Lres$ seemingly overlaps that shown in the panel
labeled CR, yet $F(\Lres)$ is almost featureless near this value
because the function is computed for a different resonance.  The DRR
feature persists for about the same duration as the CR, but is again
too weak to be visible at the last time shown.

The dominance of CR also fades in the latest stages when the bar speed
has more than halved.  A very mild flattening of $F$ at the ILR is
possibly present at $t=1000$, and is the only indication I could find
in plots of $F(\Lres)$ of changes to $F(\Lres)$ at this resonance.
However, the weakness of features in $F(\Lres)$ at this resonance does
not imply that no changes occur; indeed Fig.~\ref{fastf} shows that
substantial interactions are taking place at the ILR around this time.
The weakness of this feature at the ILR may indicate that exchanges at
this resonance are not self-reinforcing, in contrast to those at other
major resonances.  Alternatively, features may be suppressed by two
effects: the most important is that the instantaneous angular momenta
of particles that have passed through the ILR oscillate by amplitudes
of typically 0.05 (see Fig.~\ref{fastf}c), which may be sufficient to
wash out intrinsic features.  Second, conservation of $J_f$ at this
resonance implies that changes in $J_r$ must be reflected in changes
of the opposite sign in $L_z$, leading to partial cancellation in the
net change in $L_z$.

\section{Time dependence}
\label{tdep}
Weinberg (2004) finds that the LBK torque formula from TW84 does not
predict the correct time evolution of the bar pattern speed; he was
able to obtain approximate agreement with his experimental results
only by taking the history of the bar perturbation into account.  His
formula for the torque in the time-dependent regime is also of the
form eq.~(\ref{LBK}), but with a function $\Theta$ that now depends on
the entire history of the perturbation.

Since a theoretical prediction invites comparison, I have tried
experiments similar to that reported by Weinberg.  The solid curve in
Figure~\ref{turnon} shows the pattern speed evolution using my code
with the bar and NFW halo (see Appendix) that Weinberg employed for
his first test.  The axes have been scaled to the units adopted by
Weinberg and I have reproduced the curves from his figure (see also
Fig.~16 of WK05); his time-dependent, linear theory prediction is
shown by the dashed curve and Weinberg's own simulation is shown by
the dot-dashed curve.  Both curves differ from each other and from my
result, although the differences are minor.

The small differences from both Weinberg's theoretical prediction and
from his simulation do appear to be significant, however.  I have
checked that my result does not depend on any numerical parameters;
increasing the number of particles to 10M, halving the time step,
increasing the outer truncation radius by 50\%, and other tests all
yielded results that differ from the solid curve shown by scarcely
more than the thickness of the line.  I have also verified that total
angular momentum is conserved to 0.3\% of the small total possessed by
the bar at the start.

\begin{figure}[t]
\plotone{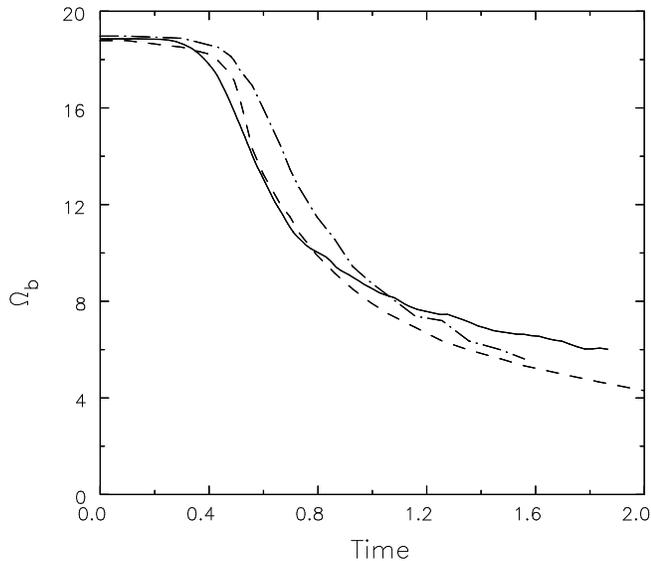}
\caption{\footnotesize The dashed line reproduces Weinberg's (2004)
prediction from perturbation theory for the pattern speed evolution,
while the dot-dash line shows his $N$-body simulation result; the
model he used is described in the Appendix.  The solid line shows the
evolution in my experiment for the same case.  See \S\ref{tdep} for
details.}
\label{turnon}
\end{figure} 

Weinberg's prediction requires sophisticated numerical treatment of a
stiff system of coupled equations, and differences from my simulation
could possibly result from numerical inadequacies in his prediction.
It is harder to understand why Weinberg's simulation differs from
mine, as I have found my result to be so robust.

Returning to Fig.~\ref{vscale}(a), the various dotted curves indicate
a strong, and long-lasting, dependence on the initial pattern speed.
I have examined orbits and $F$ for the dramatically stronger friction
exhibited by the curve shown by the solid line in this figure, in the
same manner as for the fiducial model reported in \S\ref{orbits}.  I
find that the stronger drag at a given $\Omega_b$ when the pattern
speed is started from a higher value arises from a steeper local
gradient in $F(\Lres)$ because the shoulder is more pronounced than
that shown in Fig.~\ref{Lsmooth}.  The resonance lies in similar
position relative to the peak, but the higher peak has steeper slopes.
The shoulder is more pronounced because the higher starting speed
enables more particles at higher frequencies to be caught in the CR,
and to experience much larger changes in $L_z$ (some gain by up to
factors of 5).

\subsection{Discussion of time-dependence}
In hindsight, the importance of time-dependence is entirely
reasonable.  It is clear from the distribution of particles about the
CR and OLR shown in Fig.~\ref{Lsmooth} that the local slope of $F$
depends on the past history of resonant interactions.  As it develops,
friction sculptures the \DF\ in the vicinity of the dominant
resonance, which in turn affects the strength of the subsequent
frictional drag.  Thus the limit of a steady bar rotating at fixed
$\Omega_b$, which was adopted by TW84, does not capture the true
story; Weinberg's (2004) re-derivation using Laplace transforms is
essential.

The very strong dependence of the acceleration on the starting angular
frequency reported in Fig.~\ref{vscale}(a) is reflected in the
magnitude of the shoulder, and the local gradient in $F$ at the
resonance.  The curves in that figure seem to converge for smaller
values of $\Omega_b$, suggesting somewhat less extreme dependence on
past history in the physically more reasonable regime where corotation
is beyond the end of the bar.

Figs.~\ref{mscale} -- \ref{sigmascale} show that the scaling of the
frictional force with the main parameters of eq.~(\ref{LBK}) still
holds, at least approximately, despite the complication of
time-dependence, as Weinberg also notes.  Note that the perturbation
has the same initial $\Omega_b$ and turn-on rule for the bar mass and
halo density tests, so that the time-dependence of the function
$\Theta$ did not change.  However, the initial bar pattern speed had
to be scaled in the experiments with different velocity dispersion,
making the reasonable agreement shown in Fig.~\ref{sigmascale}(b) all
the more reassuring.

\section{Convergence test}
\label{converg}
As noted earlier, halo particles gain angular momentum, on average, at
resonances.  Even though each may pass through the resonance rapidly,
we will not obtain the smooth torque expected in the continuum limit
unless there are many particles, densely spread over a broad range of
frequencies.  The required number depends on the width of the
resonance in frequency space: If the bar had a fixed or very slowly
changing pattern speed, the resonances would be sharp and the
simulations would indeed require a very large $N$ (Weinberg \& Katz
2002; \HBWK\ 2003; WK05), but the forcing frequency is broadened by
its decreasing angular speed and we can hope that the discreteness of
the particle distribution will become insignificant for some
attainable $N$.

\begin{figure}[t]
\plotone{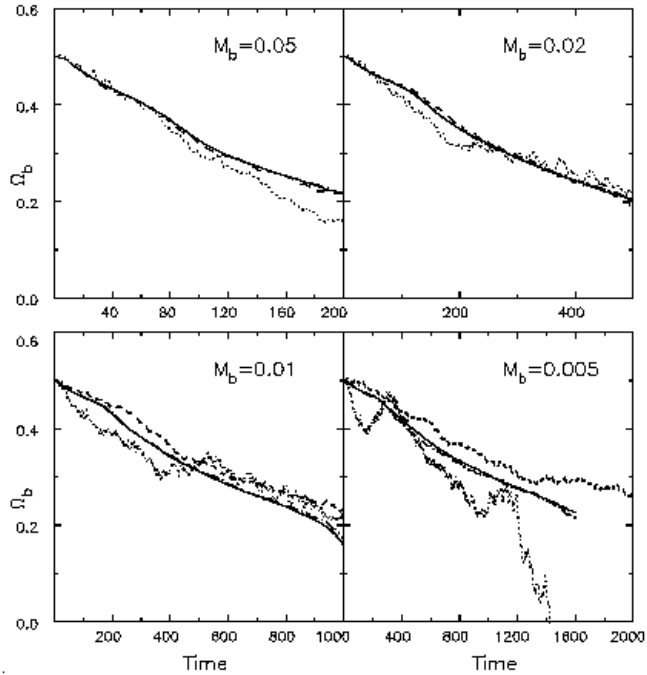}
\caption{\footnotesize The bar pattern speed as a function of time for
experiments with differing mass bars, each for four separate values of
$N$.  The indicated bar masses are fractions of $M_h$ and the length
of the time axis is inversely proportional to the bar mass in each
panel.  The line styles indicate the number of particles: $N=10^4$ --
dotted, $N=10^5$ -- dashed, $N=10^6$ -- dot-dashed, and $N=10^7$ --
solid.}
\label{Nscale}
\end{figure}

Figure \ref{Nscale} shows that the pattern speed evolution in the
Hernquist halo appears to converge to a smooth function as $N$
increases.  Merely $10^4$ particles are sufficient to obtain
qualitatively correct behavior for the 5\% mass bar, but almost one
hundred times larger $N$ is needed for comparable numerical quality
with a bar of one tenth this mass.  Phase space needs to be populated
more densely as the mass of the bar decreases, in order that a large
enough number of particles are passing through the resonance at any
time to yield a smoothly varying force.  Both the effective potential
of the resonance is weaker, and the Lorentzian frequency width of the
resonance is reduced by the slower braking rate.  As both factors
vary with the mass of the perturber, it seems reasonable that the
number of particles needed to obtain smoothly varying friction should
increase faster than $M_b^{-1}$.  However, one million is adequate
even for a low-mass bar.

The smoothness of the braking rate in Figs.~\ref{vscale} \&
\ref{msplum}, and for large $N$ in Fig.~\ref{Nscale}, is evidence that
the resonant exchanges which give rise to the force are dense enough
at most frequencies to produce a smooth torque.  The frictional force
may become erratic at very low pattern speed (Fig.~\ref{vscale}a)
because phase space is not populated densely enough as
$|\dot\Omega_b|$ decreases, but this is a physically uninteresting
regime.

These empirical results conflict with the recent claims by WK05 that a
much larger number of particles is needed to obtain the correct
result.  However, their analysis does not take proper account of the
changing pattern speed, which is so important.  The time dependence of
the frictional force arises from preceding changes to the \DF, and the
shoulders in Fig.~\ref{Lsmooth} indicate directly just how broad the
resonance is.  The feature can develop and behave as it does only if
particles with angular momenta as far away as the width of the
shoulder from the precise resonance are being affected by the
resonance.  Since $\sim 10$\% of all halo particles contribute to each
panel, it appears that $\ga1$\% of all halo particles take part in
exchanges at a dominant resonance at any moment.  The convergence
tests in Fig.~\ref{Nscale} confirm that stochastic behavior when
resonances are too sparsely populated (what WK05 describe as
``coverage'') is brought under control at comparatively low-$N$, when
the pattern speed changes at realistic rates.

It should be noted that these experiments reveal only the number of
particles needed to populate phase space densely enough for the
frictional force to approximate the continuum limit in the absence of
self-gravity.  The issue of collisional relaxation also needs to be
addressed when the particles interact with each other, which may
require more particles, as I discuss in \S\ref{ctest}.

\section{Effect of self-gravity in the halo}
\label{selfg}
All experiments reported so far treat the halo as a non-interacting
population of test particles, which move in a fixed potential well and
respond only to the perturber, as in the usual treatment of dynamical
friction (\eg\ BT).  Collective effects in a collisionless halo could
also be important, however, because the wake itself contributes to the
non-axisymmetric density affecting the orbits of the halo particles.
Hernquist \& Weinberg (1992), Weinberg \& Katz (2002), and Sellwood
(2003) have already reported some experiments with rigid bars that
take this into account, but the emphasis in those papers was on the
change to the halo density profile.

Here I study the effect of self-gravity on the frictional force, while
still employing an imposed bar.  I break the discussion of collective
effects into two levels of complication: when self-gravity includes
the monopole terms, the radial mass profile of the halo could possibly
change substantially over time, as claimed by Weinberg \& Katz (2002).
I therefore begin by including only the quadrupole field of the
response density, before describing how other terms affect the
evolution.  Note that all the experiments described in this paper are
perturbed with the non-axisymmetric field of a rigid bar.  Fully
self-consistent simulations, with the bar also made of responsive
particles, are reported in other work (\eg\ Papers II \& III).

\begin{figure}[t]
\plotone{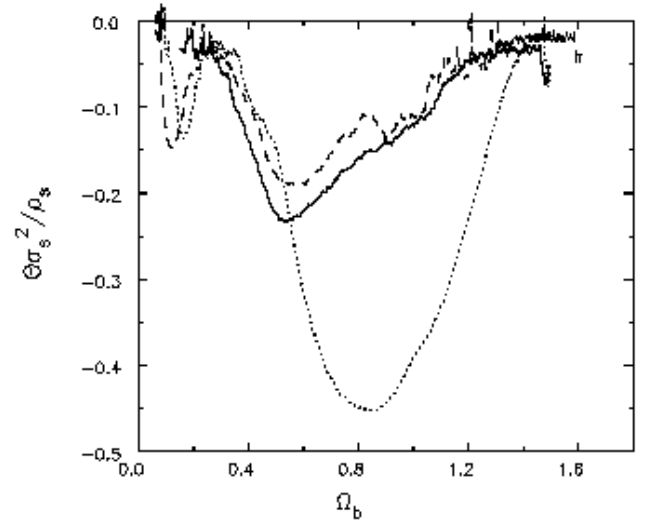}
\caption{\footnotesize The acceleration of the bar when self-gravity
of the halo response is included.  The solid curve shows the behavior
when all terms $0 \leq l \leq 4$ are included, the dashed curve shows
the situation when only the $l=2$ terms are employed.  The dotted
curve shows the corresponding case with no self-gravity, the fiducial
model, reproduced from Fig.~\ref{vscale}(a).}
\label{quadonly}
\end{figure}

I compute the self-gravity of the halo using the method described by
McGlynn (1984) and Sellwood (2003, Appendix A).  Briefly I use a 1-D
spherical grid, with an expansion of the gravitational field up to
order $l_{\rm max}$ in surface harmonics on each radial shell of the
spherical grid.

\subsection{Quadrupole field of the response}
The dashed curve in Figure \ref{quadonly} shows the bar acceleration
when only the quadrupole field of the halo response density is
included; the dotted curve, reproduced from Fig.~\ref{vscale}(a),
shows the behavior with no self-gravity.  It is clear that the
self-gravity term increases the drag slightly when the pattern speed
is slow, but greatly {\it diminishes\/} it when the bar has an
artificially high pattern speed.

This behavior is a consequence of the phase of the halo response.
Including the self-gravity of the response weakens the net torque on
the bar when the response is more than $45^\circ$ out of phase with
the imposed bar; the phase lag when self-gravity is included is
somewhat similar to that shown in Fig.~\ref{vscale}(b), but the
response remains more nearly perpendicular for longer.  As the bar
slows, the response gradually shifts towards alignment with the bar,
thereby augmenting the quadrupole field of the perturbation and
causing modestly increased friction.  Note that friction peaks at
$\Omega_b \simeq 0.5$ when corotation is at the end of the bar,
implying that self-gravity always enhances friction in the physically
relevant regime.

\subsection{Including the monopole and other terms} 
The solid line in Fig.~\ref{quadonly} shows the effect of adding more
terms to the self-gravity of the halo response; in this case all $l
\leq 4$ terms.  The addition of these extra self-gravity terms
increases friction somewhat at most angular speeds over that obtained
when only $l=2$ terms are used.

\subsection{Convergence test}
\label{ctest}
\begin{figure}[t]
\plotone{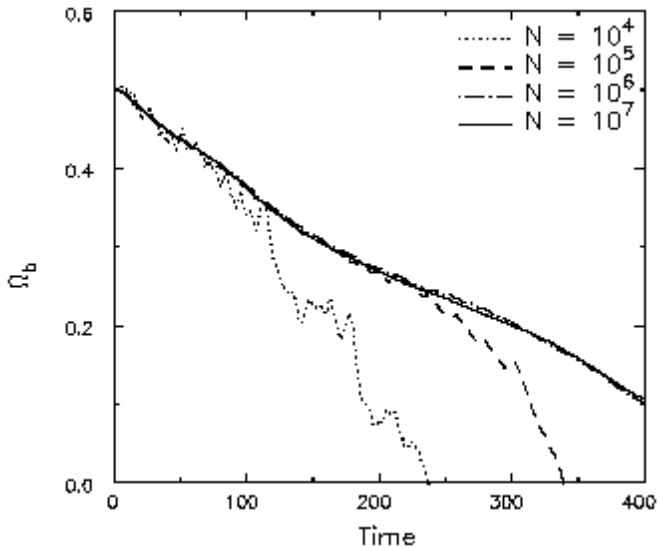}
\caption{\footnotesize The evolution of the pattern speed of a 2\%
mass bar which starts with corotation at the end of the bar ($a=r_H$),
for experiments with different $N$.  Unlike for Fig.~\ref{Nscale},
these experiments include all low-order self-gravity terms of the halo
particles ($l \leq 4$).}
\label{selfgfig}
\end{figure}

Figure~\ref{selfgfig} shows the effect of changing the particle
number, $N$.  I use the Hernquist halo, a bar with $M_b = 0.02M_h$,
and a more realistic initial $\Omega_b$.  This test includes all $l
\leq 4$ terms of the self-gravity of the halo density response and
thus some collisional relaxation must be present.

As without self-gravity, the time variation of the pattern speed
becomes smoother as $N$ increases.  The evolution is closely similar
for the two cases with $N\geq10^6$, but friction is clearly
overestimated when $N=10^4$ and slightly so when $N=10^5$.  The
minimal differences between the results for the experiments with
$N=10^6$ and $N=10^7$ indicate that 1M particles is a sufficient
number to capture the correct physics for this case with a low-mass
bar.

The 2\% mass bar used in Fig.~\ref{selfgfig} is the same as for the
second most strongly braked case in Fig.~\ref{Nscale}.  Comparison of
these two convergence tests indicates that the low-$N$ models depart
more strongly from the high-$N$ results when self-gravity is included.
However, the number of particles needed for convergence is not
increased dramatically by self-gravity.  Thus, Weinberg \& Katz (2002)
and \HBWK\ (2003) are correct that the reduction in orbit quality
caused by numerical noise in self-gravitating halos does affect the
friction force, but the torque converges for reasonably accessible
numbers of particles; 1M appears to be plenty in this problem, as
Sellwood (2003) concluded from fully self-consistent experiments.

Note also the complete absence of numerical noise in the higher-$N$
experiments shown in Figs.~\ref{Nscale} \& \ref{selfgfig}.  The
convergence is truly impressive; curves for which $N$ differs by
factors of 10 overlay almost perfectly, and there is little evidence
of differences due to shot noise.\footnote{This aspect is, in part,
due to the careful selection of particles from the \DF\ for the
initial set-up (see Debattista \& Sellwood 2000 Appendix B); random
selection of particles inevitably leaves $\sqrt N$-type variations in
the density of particles as a function of the integrals and does not,
therefore, achieve quite such impressive convergence.}  It seems
inconceivable that such impressive convergence could be achieved if
Weinberg's arguments (\eg\ Weinberg \& Katz 2002 and subsequent
papers) for the much higher number of particles needed were correct.

In this vein, it should also be noted that my earlier work to
reproduce the normal modes of disks (\eg\ Sellwood 1983; Earn \&
Sellwood 1995) already demonstrated that simulations with modest
numbers of particles could capture global instabilities, which are
driven by resonant interactions between particles and rotating
potential perturbations.  The eigenfrequency predicted from linear
theory can be reproduced to a precision of a percent or two in
simulations with a few tens of thousands of particles.  Admittedly
these results were for a dynamical instability in a thin disk, whereas
halo friction is a secular effect in 3-D.  However razor-thin disks
with random motion are scarcely more complicated than a spherical
model, where each orbit is confined to its own plane.  Furthermore,
since the evolution in both cases is determined by the resonant terms,
they have similar requirements for phase coverage of particles at the
resonance, the ability of the resonance to persist for a large number
of dynamical times, \etc, and there is no clear reason why particle
number need be orders of magnitude higher for secular evolution than
for instability.

\subsection{Claimed counter-example}
\HBWK\ (2003) report fully self-con\-sistent experiments for which the
behavior with $N=1$M differs from similar experiments with larger $N$.
However, the difference does not necessarily support their conclusion
that 1M particles is inadequate for bar-halo interaction.  They show
(their Fig.~20) that the quadrupole field in the 1M particle
experiment is weaker, so the halo torque {\it must\/} also be weaker,
which implies that the lower halo particle number could still be
adequate.  Both Weinberg \& Katz (2002) and Sellwood (2003) report
that smaller $N$ leads to more rapid angular momentum transfer, as
also shown in Fig.~\ref{selfgfig}, making the inadequate-$N$
interpretation in this case still less plausible.  \HBWK\ do indeed
have a result that is $N$-dependent, but the effect of lower particle
number is to decrease the strength of the $m=2$ distortion in the
disk, and is therefore a consequence of disk dynamics and not halo
friction.  It is likely that a higher noise level in their lower-$N$
disk interferes with their bar triggering mechanism.  Specifically,
they start with a bar-unstable disk and also apply a transient tidal
field to trigger the bar; the smaller the number of particles, the
larger the initial amplitude of the intrinsic instability, which will
generally interfere with the applied tidal field leading to a weaker
bar, unless the phases of the two bar-forming mechanisms happened to
be nearly aligned.

\section{Conclusions}
I have shown that dynamical friction between a rotating, imposed bar
and a halo scales in a very similar manner to that predicted by
Chandrasekhar's formula.  The experiments reported in
\S\S\ref{restrict} \& \ref{scalings} are highly idealized and employed
the simplest possible system that could capture dynamical friction on
a bar rotating in a halo.  The retarding acceleration depends on the
angular speed of the bar, scales linearly with its mass (\ie\ the
strength of its quadrupole field) and with the background density.  It
also scales roughly inversely as the square of the velocity dispersion
of the background through which it moves, although this scaling is
more approximate unless the model is self-similar.  This result is the
analog for a bar of that obtained by Lin \& Tremaine (1983) from a
similar study with orbiting satellites.

Even though the physical situation is quite different from rectilinear
motion through an infinite, uniform background, and complicated by the
existence of resonances and bar turn-on issues, the Chandrasekhar
scaling still holds.  It holds, because dynamical friction is
fundamentally a second order effect that arises from the interaction
of a perturber with its own wake.  The strength of the interaction
does depend on the details, but the parameter scaling cannot.

In all three halo types employed here, friction is weak when the bar
pattern speed is so high that corotation is well inside the bar.  As
the bar slows, the frictional drag grows at first but generally peaks
before corotation reaches the bar end.  Thus, in the physically
interesting regime, where corotation is beyond the end of the bar,
friction always decreases as the bar slows in these non-rotating
halos.

Friction is dominated by the quadrupole field of the bar both because
the quadrupole is the dominant potential component of the bar, but
also because the higher-order resonances that are associated with the
higher expansion terms couple less strongly to the particles.  Since
the lowest-order, non-axisymmetric component of the bar field
dominates, friction can be captured adequately in simulations with
even quite low spatial resolution, provided enough particles are
employed.

Tremaine \& Weinberg (1984) first demonstrated that friction arises as
the perturbation sweeps across resonances with the particles, a point
stressed in recent work by Weinberg and his co-workers and by
Athanassoula (2003).  I have shown that the pattern speed of the bar
changes sufficiently rapidly that a halo particle generally passes
through the resonance without any complicated non-linear trapping, as
expected by TW84.  The forced responses of the halo particles change
from anti-alignment to alignment with the bar as the pattern speed
crosses a resonance, which happens smoothly because the resonance is
broadened by the changing pattern speed.  The net effect is to produce
a global density response that lags the bar, causing the frictional
drag.  The lag angle between the bar and the halo response varies with
the friction force, as shown in Fig.~\ref{vscale}(b); it is close to
$45^\circ$ when friction is strong, but the halo response is closely
aligned with the bar when the bar is slow, and almost orthogonal when
the bar is unreasonably fast.

Since one of the two actions associated with the unperturbed motion is
adiabatically invariant, exchanges at resonances can be examined as a
function of a single angular momentum-like variable, which I denote
$\Lres$.  The density of particles is generally a decreasing function
$\Lres$, although previous angular momentum exchanges with the bar
cause a local shoulder to develop at the most important resonances.
The distribution with $\Lres$ evolves most strongly at corotation at
physically interesting pattern speeds, but smaller changes are
detectable at other resonances before the bar has slowed much.

The sculpturing of the particle distribution about the principal
resonances by previous friction seems to be the reason for the strong
dependence of friction on the previous evolution, discovered by
Weinberg (2004).  The magnitude of the friction force is determined by
the local gradient in $F$, which differs from that in the initial
model because previous exchanges with the perturbation have rearranged
the particle distribution.  A possibly similar effect may happen in
proto-planetary disks (Artymowicz 2004), although self-reinforcing
responses in that context have been found only at corotation, whereas
I have found them at other resonances also.

In the absence of self-gravity, the number of particles needed to
obtain a smoothly varying frictional force is quite modest, unless the
bar is very weak.  The main requirement here is that the broadened
resonances caused by the time-varying pattern speed should overlap
many particles in order to obtain the correct density response.  The
pattern speed changes more slowly for weaker bars, decreasing the
frequency width of the perturbation and consequently raising the
particle number needed to obtain a smoothly varying force.  The number
of particles needed in this regime appears to rise more steeply than
the inverse of the quadrupole field strength.

Weinberg \& Katz (2002) and WK05 argue that collisional relaxation in
simulations with self-gravity should increase the number of particles
required to approach the continuum limit, because potential
fluctuations arising from Poisson noise affect the orbital behavior.
The lowest-energy orbits, which they find are most delicate, should
make a negligible contribution to the torque.  I have found a small
reduction in force quality when self-gravity is included, consistent
with the prediction by these authors, but quite modest particle
numbers are needed to bring this problem under control.  Thus the
frictional torque can be simulated accurately with standard algorithms
and readily accessible computers.

All experiments reported here employ an imposed bar potential in order
to examine the dependence of the halo response on the bar parameters.
This simplifying approximation has many obvious disadvantages, as
noted in \S\ref{models}.  Fully self-consistent simulations, such as
those to be reported in later papers in this series, are the only way
to ensure that the bar forms with a dynamically realistic amplitude
and pattern speed and responds to friction in the appropriate way.

\acknowledgments I am grateful to Martin Weinberg for very helpful
discussions, running some tests for comparison, sending a draft of his
preprint, and for comments on an earlier draft of this paper.  I also
thank Victor Debattista for many thoughtful suggestions, Kelly
Holley-Bockelmann, Tad Pryor, and Juntai Shen for input, and Scott
Tremaine for some helpful comments.  An anonymous referee drew my
attention to Artymowicz's paper.  This work was begun while the author
was a visiting member of the Institute for Advanced Study in
Princeton; their hospitality is gratefully acknowledged.  Grants from
NASA (NAG 5-10110) and from NSF (AST-0098282) provided support.

\appendix
%\section*{Appendix}
\bigskip
The NFW (Navarro, Frenk \& White 1997) halo density profile is
\begin{equation}
\rho_{\rm NFW}=\frac{\rho_s r_s^3}{r(r+r_s)^2},
\end{equation}
where $\rho_s$ is a scale density and $r_s$ is a scale length.  The
concentration parameter $c$ is defined as the ratio of the ``virial
radius'' to $r_s$; at the virial radius, the {\it average\/} enclosed
density is $200 \rho_{\rm crit} = 200 \times 3H_0^2/(8\pi G)$, with
$H_0$ being Hubble's constant.  Eddington's formula (Binney \&
Tremaine 1987, eq.~4-140b) yields an isotropic distribution function
that is positive everywhere, but which requires some care to evaluate
for the most bound energies.

Note that the NFW profile has a more slowly declining outer density
gradient than does the Hernquist profile (eq.~\ref{Hernquist}), but is
otherwise closely similar.  The density and potential when $r<r_s$ are
almost the same, while the logarithmic mass divergence of NFW profiles
is both numerically inconvenient and physically unimportant -- the
extra mass at large radii has such low orbital frequencies and feels
such a weak perturbing field from the distant quadrupole that its
contribution to the total torque is negligible at pattern speeds of
interest.

Weinberg (2004) uses an NFW halo with $c=15$ for his first application
of his formulae.  I adopt his units for this case to facilitate
comparison with his result; he chooses $15r_s$ to be his length unit,
the mass interior to this radius as his mass unit, and he also sets
Newton's constant $G=1$.  He chose a large, strong, heavy bar, with
semi-major axis equal to $r_s = 1/15$, axis ratios $a:b:c =
1:0.2:0.05$, and a mass equal to half the halo mass interior to this
radius, or 0.0526 of the virial mass.  The initial pattern speed of
the bar is 18.84 in his units, which places corotation at the bar end.
The bar amplitude varies as $f(t) = [1 + {\rm erf} (4t - 2)]/2$, where
$t$ is the time in Gyr for a halo scaled to the Milky Way; when $r_s =
20\;$kpc, Weinberg's time unit is $\sim 1.4\;$Gyr.

\end{document}